\documentstyle[multicol,aps,epsf]{revtex}

\title{Molecular dynamics simulations of spin and pure
       liquids with preserving all the conservation laws}

\author{I. P. Omelyan,$^1$ I. M. Mryglod,$^{1,2}$ and R. Folk$^2$}

\address{$^1$Institute for Condensed Matter Physics,
         1 Svientsitsky Street, UA-79011 Lviv, Ukraine}
\address{$^2$Institute for Theoretical Physics, Linz University,
         A-4040 Linz, Austria}

\date{\today}

\begin{document}

\maketitle

\begin{abstract}

A new methodology is developed to integrate numerically the equations of
motion for classical many-body systems in molecular dynamics simulations.
Its distinguishable feature is the possibility to preserve, independently
on the size of the time step, all the conservation laws inherent in the
description without breaking the time reversibility. As a result, an
implicit second-order algorithm is derived and applied to pure as well
as spin liquids for which the dynamics is characterized by the conservation
of total energy, linear and angular momenta as well as magnetization and
individual spin lengths. It is demonstrated on the basis of Lennard-Jones
and Heisenberg fluid models that when such quantities as energy and
magnetization must be conserved perfectly, the new algorithm turns out
to be more efficient than popular decomposition integrators and standard
predictor-corrector schemes.

\vspace{12pt}

\noindent
Pacs numbers: 02.60.Cb; 75.40.Gb; 75.50.Mm; 76.50.+g; 95.75.Pq

\end{abstract}

\vspace{16pt}

\begin{multicols}{2}

\section{Introduction}

During the last years, considerable attention has been focussed on computer
experiment studies of relaxation properties and critical phenomena in
classical spin systems \cite{ChenLan,BunChen,EvLan,CosLan,Krech,LanKre,%
Landau,Tsai,Omfe}. These studies dealt mainly with lattice models such
as the Ising, the XY, and the Heisenberg. Of notable current interest is
the investigation of continuum spin liquid models \cite{Lom,Tavar,Mryglod,%
Nijmeis,Nijmei,MryglodFolk} in which additional dynamical effects are
possible because of the coupling between spin and liquid subsystems.

Quite recently \cite{Omfec,Omfes}, a set of symplectic algorithms of
different orders in the time step has been constructed for numerical
integration of motion at the presence of both translational and spin
degrees of freedom. As a consequence, the molecular dynamics simulations
(MD) of a Heisenberg spin fluid have been carried out for the first time.
The symplectic integrators were derived by developing the Suzuki-Trotter
technique \cite{Suzuki} for decompositions of exponential operators. Their
main advantages over standard predictor-corrector schemes are explicitness,
time reversibility and exact conservation of spin lengths. It was also
shown that the decomposition algorithms permit significantly larger time
steps and lead to a substantial speedup of the calculations. In a particular
case when the spin degrees of freedom are frozen, these algorithms can be
reduced to the well-known velocity Verlet integrator \cite{Swope}, widely
used for simulating of pure liquid dynamics.

However, the decomposition approach (like predictor-corrector and other
existing traditional numerical schemes \cite{Burden}, such as Runge-Kutta,
etc.) do not preserve the total energy and magnetization of the system. In
most MD applications the accuracy achieved for the energy-magnetization
conservation by the decomposition algorithms is high enough in order to
obtain reliable results. Moreover, this accuracy can be improved using
higher-order versions \cite{Omfec} of the decomposition approach or
decreasing the step size. But if the integrals of motion have to be conserved
perfectly, the non-conservative algorithms may not be an optimal choice for
the solution of the problem. The reason is that then the time step should
be divided into a lot of subintervals, reducing considerably the efficiency
of the computations.

The exact conservation of integrals of motion is especially important
for simulations of spin liquids near phase transitions, when the phase
diagrams, dynamical scaling, long-time correlated behavior or derivatives
of the thermodynamic functions are investigated. This is so because
in these cases, the presence of artificial fluctuations in energy and
magnetization may have a significant influence on the results. Therefore,
it is desirable to look for an algorithm which conserves the fundamental
physical invariants exactly or at least within machine accuracy.

In the present paper, a novel approach to numerical integration of the
equations of motion for spin and pure liquids is introduced. The main
feature of this approach is its intrinsic preservation of all the
conservation laws inherent in the system without violating the time
reversibility property. The paper is organized as follows. The basic
equations and their integrals of motion are described in Section 2.
Section 3 is devoted to a consequent derivation of the desired
second-order algorithm. Its possibility to conserve exactly the
integrals of motion is demonstrated there also. In Sections 4 and 5,
the algorithm is tested in actual MD simulations on Heisenberg and
Lennard-Jones fluid models, respectively, and compared with previous
numerical schemes. The discussion and concluding remarks are given
in Section 6.

\section{Basic equations of motion and conservation laws}

Let us consider a classical $N$-body system described by the Hamiltonian
\begin{equation}
H = \sum_{i=1}^N \frac{m_i {{\bf v}_i}^2}{2} + \frac12 \sum_{i \ne j}^N
\Big( \varphi(r_{ij}) - J(r_{ij}) \, {\bf s}_i {\mbox{\boldmath
$\cdot$}} {\bf s}_j \Big) ,
\end{equation}
where ${\bf r}_i$ and ${\bf v}_i$ are the translational position and
velocity, respectively, of particle $i$ with mass $m_i$ carrying spin
${\bf s}_i$. The fluid part of the potential is denoted by $\varphi
(r_{ij})$, whereas $J(r_{ij})$ is the exchange integral corresponding
to a pair of spins with the interparticle separation $r_{ij}=|{\bf r}_i-
{\bf r}_j|$. Note that within the classical approach, each spin ${\bf
s}_i$ is treated as a continuous three-component vector with fixed
length (putting for convenience $|{\bf s}_i| = 1$, so that $J$ will be
measured in energy units). Although the results which will be obtained
below can easily be adapted to a larger class of Hamiltonians (to
multi-component systems, for instance), we restrict ourselves for the
sake of simplicity to the basic model (1) which represents a typical
isotropic Heisenberg spin fluid \cite{Mryglod,Nijmei}. For $J \equiv
0$, Eq.~(1) reduces to a pure liquid model.

In MD simulations it is necessary to solve numerically the equations of
motion, ${\rm d} {\mbox{\boldmath $\rho$}}/{\rm d} t = [{\mbox{\boldmath
$\rho$}}, H]$, where $[\,,\,]$ denotes the Poisson bracket and ${\mbox
{\boldmath $\rho$}} \equiv \{ {\bf r}_i, {\bf v}_i, {\bf s}_i \}$ is
the full set of microscopic phase variables. For the system under
consideration, the dynamical equations can be written more explicitly
\cite{Mryglod,Omfes},
\begin{eqnarray}
\frac{{\rm d}{\bf r}_i}{{\rm d} t} &=& {\bf v}_i , \nonumber \\
\frac{{\rm d}{\bf v}_i}{{\rm d} t} &=& \frac{{\bf f}_i}{m_i} \equiv
- \frac{1}{m_i} \sum_{j (j \ne i)}^N
\! \bigg( \frac{\partial \varphi_{ij}}{\partial r_{ij}}-\frac{\partial
J_{ij}}{\partial r_{ij}} \, {\bf s}_i {\mbox{\boldmath $\cdot$}}{\bf s}_j
\bigg) \frac{{\bf r}_{ij}}{r_{ij}} , \\
\frac{{\rm d}{\bf s}_i}{{\rm d} t} &=&
\frac{{\bf s}_i}{\hbar} {\mbox{\boldmath $\times$}} {\bf g}_i \equiv
\frac{{\bf s}_i}{\hbar} {\mbox{\boldmath $\times$}} \! \sum_{j (j \ne i)}^N
J_{ij} \, {\bf s}_j . \nonumber
\end{eqnarray}
Here, ${\bf f}_i = \sum_{j (j \ne i)} {\bf f}_{ij}$ and ${\bf g}_i =
\sum_{j (j \ne i)} {\bf g}_{ij}$ are the force and internal magnetic field,
respectively, acting on particle $i$ due to the interactions ${\bf f}_{ij}=
-(\varphi'_{ij}-J'_{ij} \, {\bf s}_i {\mbox{\boldmath $\cdot$}}{\bf
s}_j) {\bf r}_{ij}/r_{ij}$ and ${\bf g}_{ij}=J_{ij} \, {\bf s}_j$ with
all the rest of bodies, where $\varphi_{ij} \equiv \varphi(r_{ij})$ and
$J_{ij} \equiv J(r_{ij})$. Note that the quantum Poisson bracket was
applied \cite{Krech,Mryglod} to derive the equations for spin subdynamics.
If an initial state ${\mbox{\boldmath $\rho$}}(0)$ is specified, the time
evolution ${\mbox{\boldmath $\rho$}}(t)$ can be uniquely obtained by
integrating Eq.~(2).

Taking into account the symmetry $\varphi_{ij}=\varphi_{ji}$ and $J_{ij}=
J_{ji}$ of interaction potentials, it follows from Eq.~(2) that the total
energy $E \equiv H$, the total magnetization ${\bf M}=\sum_i {\bf s}_i$,
the total linear momentum ${\bf P}=\sum_i m_i {\bf v}_i$
as well as angular momentum ${\bf L}=\sum_i m_i {\bf r}_i
{\mbox{\boldmath $\times$}} {\bf v}_i$ are
integrals of motion, i.e., ${\rm d} E/{\rm d} t = {\rm d} {\bf M}/{\rm d} t
= {\rm d} {\bf P}/{\rm d} t = {\rm d} {\bf L}/{\rm d} t = 0$. The structure
of spin equations of motion (the last line of Eq.~(2)) imposes in addition
the conservation of individual spin lengths, $|{\bf s}_i|=s_i=const$.
Indeed, ${\rm d} s_i/{\rm d} t = {\rm d} ({\bf s}_i {\mbox{\boldmath
$\cdot$}} {\bf s}_i)^{1/2}/{\rm d} t= ({\bf s}_i {\mbox{\boldmath $\cdot$}}
{\rm d} {\bf s}_i /{\rm d} t)/s_i \equiv {\bf s}_i {\mbox{\boldmath $\cdot$}}
[{\bf s}_i {\mbox{\boldmath $\times$}} {\bf g}_i]/\hbar s_i=0$, because
the equality ${\bf a} {\mbox{\boldmath $\cdot$}} [{\bf a} {\mbox{\boldmath
$\times$}} {\bf b}] = 0$ is valid for arbitrary vectors ${\bf a}$ and ${\bf
b}$. Besides, the exact solutions are time reversible, since the equations
of motion are invariant with respect to the time inversion transformation
$t \rightarrow -t$, $\{ {\bf v}_i, {\bf s}_i \} \rightarrow \{ -{\bf v}_i,
-{\bf s}_i \}$.

No existing numerical scheme can obey perfectly all the just mentioned
properties. The exact conservation during the integration can be achieved
only for some of the integrals of motion, such as linear momentum, for
example. Usually, it is required that the deviations of conservative
quantities from their original values to be within an acceptable level
of precision. This results, however, in limitations on the size of the
time steps which actually can be used for MD simulating.

\section{The method of integration}

We will show in this section that it is possible to generate time-reversible
microscopic trajectories along of which all the integrals of motion are
preserved at arbitrary finite time steps. Our derivation of the desired
algorithm is started by considering a mid-point scheme of the second
order. According to this scheme, the dynamical variables can be
propagated as
$$
{\mbox{\boldmath $\rho$}}(t+\tau)={\mbox{\boldmath $\rho$}}(t) + \tau
[{\rm d} {\mbox{\boldmath $\rho$}}/{\rm d} t]_{t+\tau/2} + {\cal O}(\tau^3)
$$
with $\tau$ being the step size and ${\cal O}(\tau^3)$ denoting the
truncation terms. In view of Eq.~(2), the explicit expressions for such
a propagation read
\begin{eqnarray}
{\bf r}_i(t+\tau)&=&{\bf r}_i(t) +
\tau {\bf v}_i(t+{\textstyle \frac{\tau}{2}}), \nonumber \\
{\bf v}_i(t+\tau)&=&{\bf v}_i(t) +
\tau {\bf f}_i(t+{\textstyle \frac{\tau}{2}})/m_i , \\
{\bf s}_i(t+\tau)&=&{\bf s}_i(t) + \tau [{\bf s}_i {\mbox{\boldmath
$\times$}} {\bf g}_i]_{t+\frac{\tau}{2}}/\hbar , \nonumber
\end{eqnarray}
where the mid-step values of ${\bf v}_i$, ${\bf f}_i$ and ${\bf s}_i
{\mbox{\boldmath $\times$}} {\bf g}_i$ should be specified additionally.

The only way to construct mid-point translational velocities maintaining
the time reversibility property is
\begin{equation}
\tau {\bf v}_i(t+{\textstyle \frac{\tau}{2}}) = \frac{\tau}{2} \Big[
{\bf v}_i(t)+{\bf v}_i(t+\tau) \Big] + {\cal O}(\tau^3) .
\end{equation}
The terms ${\cal O}(\tau^3)$ of third- and higher-orders can be ignored
because the corresponding terms of the same orders have been truncated
already within the mid-point approach. Eq.~(4) represents, in fact, an
implicit interpolation formula in which past (at time $t$), and future
(at $t+\tau$) values of dynamical quantities enter symmetrically,
assuring automatically the reversibility of the solutions.

In the case of translational forces, there are several possibilities
to build the mid-point values. The reason is that the interparticle
function ${\bf f}_{ij} \equiv {\bf f} ({\bf r}_{ij}, {\bf s}_i {\mbox
{\boldmath $\cdot$}} {\bf s}_j)$ depends explicitly on relative position
${\bf r}_{ij}$ and orientation ${\bf s}_i {\mbox{\boldmath $\cdot$}}
{\bf s}_j$ which in turn vary with time. Thus, we can apply the mid-point
interpolation either to the function ${\bf f}_{ij}$ as a whole, or directly
to the dynamical variables ${\bf r}_{ij}$ and ${\bf s}_i {\mbox{\boldmath
$\cdot$}} {\bf s}_j$. As a result, two different approaches to the force
evaluation may be introduced, namely,
$$
{\bf f}_{ij}(t+{\textstyle \frac{\tau}{2}}) = \frac12 \Big[{\bf f}\big(
{\bf r}_{ij}(t), [{\bf s}_i {\mbox{\boldmath $\cdot$}} {\bf s}_j]_t\big)
+ {\bf f}\big({\bf r}_{ij}(t+\tau),[{\bf s}_i {\mbox{\boldmath $\cdot$}}
{\bf s}_j]_{t+\tau}\big) \Big]
$$
and
\begin{equation}
{\bf f}_{ij}(t+{\textstyle \frac{\tau}{2}}) = {\bf f}\big({\bf r}_{ij}(t+
{\textstyle \frac{\tau}{2}}), [{\bf s}_i {\mbox {\boldmath $\cdot$}}
{\bf s}_j]_{t+\frac{\tau}{2}} \big) .
\end{equation}
The last approach requires the knowledge of mid-point values for ${\bf
r}_{ij}$ and ${\bf s}_i {\mbox{\boldmath $\cdot$}} {\bf s}_j$. The
obvious choice for the relative positions is
\begin{equation}
{\bf r}_{ij}(t+{\textstyle \frac{\tau}{2}}) = \frac12 \Big[
{\bf r}_{ij}(t)+{\bf r}_{ij}(t+\tau) \Big] .
\end{equation}
The interpolation of the scalar product $[{\bf s}_i {\mbox{\boldmath
$\cdot$}} {\bf s}_j]_{t+\tau/2}$ will be described latter. None of the
above approaches for evaluating ${\bf f}_{ij}(t+\tau/2)$ can lead to
a scheme with exact preservation of the total energy. The energy will
only be conserved approximately with the precision within which the
microscopic solutions are calculated, i.e., $E(t+\tau)=E(t)+{\cal O}
(\tau^3)$.

We will show now that the second approach (Eq.~(5)) can be modified
in such a way as to compensate the loss of precision in the total energy.
The idea lies in the following. Since, according to Eq.~(1), the energy
difference $E(t+\tau)-E(t)$ is a function of the quantities $\varphi(r_{ij}
(t+\tau))$ and $\varphi(r_{ij}(t))$ as well as $J(r_{ij}(t+\tau))$ and
$J(r_{ij}(t))$, it is natural to try to evaluate numerically the partial
derivatives $\varphi'(r_{ij})=\partial \varphi/\partial r_{ij}$ and $J'
(r_{ij})=\partial J/\partial r_{ij}$ (which appear in Eq.~(5) at $t+
\tau/2$) in terms of the same quantities, rather than to calculate the
derivatives analytically. This is possible because for any function
$\xi(r_{ij})$ depending only on the interparticle distance $r_{ij}$
we can write the following two expressions
$$
\frac{{\rm d} \xi(r_{ij})}{{\rm d} t} = \frac{\partial \xi}{\partial
{\bf r}_{ij}} \frac{{\rm d} {\bf r}_{ij}}{{\rm d} t} = \xi'(r_{ij})
\frac{{\bf r}_{ij} {\mbox{\boldmath $\cdot$}} {\bf v}_{ij}}{r_{ij}}
$$
and
$$
\frac{{\rm d} \xi(r_{ij})}{{\rm d} t} \bigg|_{t+\frac{\tau}{2}} =
\frac{\xi\big(r_{ij}(t+\tau)\big)-\xi\big(r_{ij}(t)\big)}{\tau} +
{\cal O}(\tau^2) ,
$$
combining of which gives
\begin{equation}
\frac{\xi'(r_{ij})}{r_{ij}} \bigg|_{t+\frac{\tau}{2}} =
\frac{\xi\big(r_{ij}(t+\tau)\big)-\xi\big(r_{ij}(t)\big)}
{\tau {\bf r}_{ij}(t+\frac{\tau}{2}) {\mbox{\boldmath $\cdot$}}
{\bf v}_{ij}(t+\frac{\tau}{2})} + {\cal O}(\tau^2) ,
\end{equation}
where the mid-point values of relative velocity ${\bf v}_{ij}={\bf v}_i-
{\bf v}_j$ are calculated according to Eq.~(4) as
$$
{\bf v}_{ij}(t+{\textstyle \frac{\tau}{2}}) = \frac12 \Big[
{\bf v}_{ij}(t)+{\bf v}_{ij}(t+\tau) \Big] .
$$
Then choosing $\xi \equiv \varphi, J$, one finds the expression
\begin{eqnarray}
&& \tau {\bf f}_i(t+{\textstyle \frac{\tau}{2}}) = - \sum_{j (j \ne i)}
\frac{{\bf r}_{ij}(t+{\textstyle \frac{\tau}{2}})}{{\bf r}_{ij}(t+
{\textstyle \frac{\tau}{2}}) {\mbox{\boldmath $\cdot$}} {\bf v}_{ij}
(t+{\textstyle \frac{\tau}{2}})} \bigg( \varphi\big(r_{ij}(t+\tau)\big)
\nonumber \\
&& -\varphi\big(r_{ij}(t)\big) -\big[J\big(r_{ij}(t+\tau)\big)-
J\big(r_{ij}(t)\big)\big] [{\bf s}_i {\mbox{\boldmath $\cdot$}}
{\bf s}_j]_{t+\frac{\tau}{2}} \bigg)
\end{eqnarray}
for mid-step translational forces, where the ${\cal O}(\tau^3)$ terms
have been neglected.

Performing scalar multiplication of Eq.~(8) with the vector ${\bf v}_i(t+
\tau/2)$, then taking the sum over all the particles ($i=1,2,\ldots,N$),
and using the fact that the double sum obtained in the right-hand side
is invariant with respect to the replacements $i \leftrightarrow j$, it
can be shown that
\begin{eqnarray*}
&& \tau \sum_{i=1}^{N} {\bf f}_i(t+{\textstyle \frac{\tau}{2}})
{\mbox{\boldmath $\cdot$}} {\bf v}_i(t+{\textstyle \frac{\tau}{2}}) =
- \frac12 \sum_{i \ne j}^N \bigg( \varphi\big(r_{ij}(t+\tau)\big) \\
&& -\varphi\big(r_{ij}(t)\big)-\big[J\big(r_{ij}(t+\tau)\big)
-J\big(r_{ij}(t)\big)\big] [{\bf s}_i {\mbox{\boldmath $\cdot$}}
{\bf s}_j]_{t+\frac{\tau}{2}} \bigg) .
\end{eqnarray*}
Assuming for the moment that spin degrees of freedom are frozen (i.e., that
$[{\bf s}_i {\mbox{\boldmath $\cdot$}} {\bf s}_j]$ do not depend on time),
the last relation can be presented in the form
$$
\tau \sum_{i=1}^{N} {\bf f}_i(t+\tau/2) {\mbox{\boldmath
$\cdot$}} {\bf v}_i(t+\tau/2)=U(t)-U(t+\tau) ,
$$
where $U$ denotes the potential energy of the system. On the other hand,
multiplying the second line of Eq.~(3) by ${\bf v}_i(t+\tau/2)$ and
summating over the particles yields
\begin{eqnarray*}
&& \sum_i \frac{m_i}{2} \big({\bf v}_i(t+\tau)-{\bf v}_i(t)\big)
{\mbox{\boldmath $\cdot$}} \big({\bf v}_i(t+\tau)+{\bf v}_i(t)\big) \\
&& \equiv K(t+\tau)-K(t)=\tau
\sum_{i=1}^{N} {\bf f}_i(t+{\textstyle \frac{\tau}{2}})
{\mbox{\boldmath $\cdot$}} {\bf v}_i(t+{\textstyle \frac{\tau}{2}}) ,
\end{eqnarray*}
where $K$ denotes the kinetic energy. We see, therefore, that during the
time propagation given by Eqs.~(3) and (8), the total energy $E=K+U$ is
conserved exactly for any $\tau$, i.e., $E(t+\tau)=E(t)$.

Note that Eqs.~(7) and (8) are well defined when the scalar product ${\bf
r}_{ij}(t+\tau/2) {\mbox{\boldmath $\cdot$}} {\bf v}_{ij}(t+\tau/2)$ tends
to zero. This is so because according to the first line of Eq.~(3), the
mid-step relative velocity is connected with the change in position by the
constraint ${\bf v}_{ij}(t+\tau/2) = ({\bf r}_{ij}(t+\tau)-{\bf r}_{ij}(t))
/\tau$. So that the scalar product is merely equal to $({\bf r}_{ij}(t+
\tau)+{\bf r}_{ij}(t)) {\mbox {\boldmath $\cdot$}} ({\bf r}_{ij}(t+\tau)-
{\bf r}_{ij}(t))/(2 \tau) \equiv (r_{ij}(t+\tau)+r_{ij}(t))(r_{ij}(t+\tau)-
r_{ij}(t))/(2 \tau)$. As a result, the right-hand side of Eq.~(7) can be
rewritten in the following mathematically equivalent form
\begin{equation}
\frac{\xi\big(r_{ij}(t+\tau)\big)
-\xi\big(r_{ij}(t)\big)}{r_{ij}(t+\tau)-r_{ij}(t)}
\frac{2}{r_{ij}(t)+r_{ij}(t+\tau)} + {\cal O}(\tau^2)
\end{equation}
which reduces to $\xi'(r_{ij}(t+\tau/2))/r_{ij}(t+\tau/2) + \epsilon^2
{\cal O}(\tau^2)$ when the value $|r_{ij}(t+\tau) - r_{ij}(t)| < \epsilon$
is small enough, where $r_{ij}(t+\tau/2)=(r_{ij}(t)+r_{ij}(t+\tau))/2$ and
$\epsilon^2$ denotes a machine zero.

The exact energy conservation can also be achieved in the presence of
spin subdynamics. In order to show this, unfreeze now the spin variables,
and consider first the question of how to interpolate the vector product
${\bf s}_i {\mbox {\boldmath $\times$}} {\bf g}_i$ arising in the third
line of Eq.~(3). Again, since this product depends on time implicitly via
dynamical variables ${\bf s}_i$ and ${\bf g}_i$, we will have here a lot
of possibilities. The first of them
$$
[{\bf s}_i {\mbox{\boldmath $\times$}} {\bf g}_i]_{t+\frac{\tau}{2}} =
\frac12 \Big[ {\bf s}_i(t) {\mbox{\boldmath $\times$}} {\bf g}_i(t) +
{\bf s}_i(t+\tau) {\mbox{\boldmath $\times$}} {\bf g}_i(t+\tau) \Big]
$$
is not suitable because it does not lead to the conservation of individual
spin lengths, i.e., $s_i(t+\tau)=s_i(t)+{\cal O}(\tau^3)$. At the same time,
the second interpolation
\begin{equation}
[{\bf s}_i {\mbox{\boldmath $\times$}} {\bf g}_i]_{t+\frac{\tau}{2}} =
\frac12 \Big[ {\bf s}_i(t) + {\bf s}_i(t+\tau) \Big]
{\mbox{\boldmath $\times$}} {\bf g}_i(t+{\textstyle \frac{\tau}{2}})
\end{equation}
does conserve spin lengths exactly, $s_i(t+\tau)=s_i(t)$, for arbitrary
choice of ${\bf g}_i(t+\tau/2)$. Indeed, substituting Eq.~(10) into the
propagation equation ${\bf s}_i(t+\tau)={\bf s}_i(t) + \tau [{\bf s}_i
{\mbox{\boldmath $\times$}} {\bf g}_i]_{t+\tau/2}/\hbar$ and solving
analytically the obtained expression with respect to ${\bf s}_i(t+\tau)$,
one obtains
\begin{eqnarray}
&&{\bf s}_i(t+\tau) = \frac{1}{1+\frac{\tau^2}{4 \hbar^2}
[{\bf g}_i]^2_{t+\frac{\tau}{2}}} \Big[ {\bf s}_i(t) +
\frac{\tau}{\hbar} {\bf s}_i(t) {\mbox{\boldmath $\times$}}
[{\bf g}_i]_{t+\frac{\tau}{2}} +
\nonumber  \\
&& \frac{\tau^2}{4 \hbar^2} \Big( 2 \, [{\bf g}_i]_{t+\frac{\tau}{2}}
\big( [{\bf g}_i]_{t+\frac{\tau}{2}} {\mbox{\boldmath $\cdot$}} \,
{\bf s}_i(t) \big) - \big( [{\bf g}_i]_{t+\frac{\tau}{2}} \big)^2
{\bf s}_i(t) \big) \Big) \Big] .
\end{eqnarray}
As can be verified readily, Eq.~(11) represents an unitary transformation,
${\bf s}_i(t+\tau)={\bf \Theta}_i(t,\tau) {\bf s}_i(t)$, where ${\bf
\Theta}_i(t,\tau)$ is a rotation matrix which, of course, does not
change the norm of vectors.

Three different time-reversible interpolations can be introduced for
the factor $[{\bf g}_i]_{t+\tau/2} \equiv {\bf g}_i(t+\tau/2) = \sum_{j
(j \ne i)} [{\bf g}_{ij}]_{t+\tau/2}$. They are:
$$
[{\bf g}_{ij}]_{t+\frac{\tau}{2}} =
\frac{J\big(r_{ij}(t)\big) {\bf s}_j(t) +
J\big(r_{ij}(t+\tau)\big) {\bf s}_j(t+\tau)}{2} ,
$$
$$
[{\bf g}_{ij}]_{t+\frac{\tau}{2}} =
J \Big(\frac{r_{ij}(t)+r_{ij}(t+\tau)}{2} \Big)
\frac{{\bf s}_j(t) + {\bf s}_j(t+\tau)}{2} ,
$$
and
\begin{equation}
[{\bf g}_{ij}]_{t+\frac{\tau}{2}} =
\frac{J\big(r_{ij}(t)\big) + J\big(r_{ij}(t+\tau)\big)}{2}
\frac{{\bf s}_j(t) + {\bf s}_j(t+\tau)}{2} .
\end{equation}
The first approximation cannot be chosen for our purpose because it destroys
the total magnetization of the system, i.e., ${\bf M}(t+\tau)={\bf M}(t)+
{\cal O}(\tau^3)$. The last two interpolations do reproduce the magnetization
vector perfectly. Indeed, summing up the individual spin propagation (third
of Eqs. (3)) over all the particles and taking into account Eq.~(10) gives
${\bf M}(t+\tau)={\bf M}(t)+\Delta{\bf M}$, where
$$
\Delta {\bf M} = \frac{\tau}{4 \hbar} \sum_{i \ne j} J_{ij}(t+{\textstyle
\frac{\tau}{2}}) [{\bf s}_i(t) + {\bf s}_i(t+\tau)] {\mbox{\boldmath
$\times$}} [{\bf s}_j(t) + {\bf s}_j(t+\tau)]
$$
and $J_{ij}(t+\tau/2)$ may be equal either to $J((r_{ij}(t)+r_{ij}(t+
\tau))/2)$ or $(J(r_{ij}(t)) + J(r_{ij}(t+\tau)))/2$. The term $\Delta
{\bf M}$ is canceled because of the invariance of the double sum with
respect to the transformation $i \leftrightarrow j$, and of the obvious
equality ${\bf a} {\mbox{\boldmath $\times$}} {\bf b} + {\bf b} {\mbox
{\boldmath $\times$}} {\bf a}=0$ which fulfills for any vectors ${\bf
a}$ and ${\bf b}$. Thus, ${\bf M}(t+\tau)={\bf M}(t)$ in both the cases.
However, in the first of them when $J_{ij}(t+\tau/2)=J((r_{ij}(t)+r_{ij}
(t+\tau))/2)$, the energy difference $E(t+\tau)-E(t)$, being a function of
two quantities $J(r_{ij}(t+\tau))$ and $J(r_{ij}(t))$, cannot be reduced
to zero exactly using only one mid-point value $J((r_{ij}(t)+r_{ij}(t+
\tau))/2)$.

At the same time, within the last third interpolation given by Eq.~(12)
we are able to perform such a reduction. To demonstrate this, let us
consider finally the interpolation of the spin scalar product $[{\bf s}_i
{\mbox{\boldmath $\cdot$}} {\bf s}_j]_{t+\tau/2}$ appearing in Eq.~(8).
Similarly to Eq.~(6), one defines this interpolation in the form
\begin{equation}
[{\bf s}_i {\mbox{\boldmath $\cdot$}} {\bf s}_j]_{t+\frac{\tau}{2}} =
\frac12 \Big[ {\bf s}_i(t) {\mbox {\boldmath $\cdot$}} {\bf s}_j(t) +
{\bf s}_i(t+\tau) {\mbox {\boldmath $\cdot$}} {\bf s}_j(t+\tau) \Big] .
\end{equation}
Then, using Eqs.~(3), (8), (10) and (12), it can be shown that the
following equality holds
\begin{eqnarray*}
&& \sum_{i \ne j}^N \big[ J\big(r_{ij}(t+\tau)\big)-J\big(r_{ij}(t)\big)
\big] [{\bf s}_i {\mbox{\boldmath $\cdot$}} {\bf s}_j]_{t+\frac{\tau}{2}}
\\ = && \sum_{i \ne j}^N
\Big( J\big(r_{ij}(t+\tau)\big) {\bf s}_i(t+\tau)
{\mbox{\boldmath $\cdot$}} {\bf s}_j(t+\tau)-J\big(r_{ij}(t)\big)
{\bf s}_i(t) {\mbox{\boldmath $\cdot$}} {\bf s}_j(t)\Big) \\
\equiv && U^{\rm (s)}(t)-U^{\rm (s)}(t+\tau) ,
\end{eqnarray*}
where $U^{\rm (s)}$ denotes the spin part of the potential energy. So that,
as in the case of frozen spin subdynamics, the sum $\tau \sum_{i=1}^{N}
{\bf f}_i(t+\tau/2) {\mbox{\boldmath $\cdot$}} {\bf v}_i(t+\tau/2)$ is
reduced to the potential energy difference $U(t)-U(t+\tau)$. But as
was shown earlier using Eq.~(3), this sum can be expressed also as the
difference $K(t+\tau)-K(t)$ in the kinetic energy. This indicates again
that the total energy $E=K+U$ is conserved exactly, i.e., $E(t+\tau)
=E(t)$, despite the microscopic solutions ${\bf r}_i(t+\tau)$, ${\bf v}_i
(t+\tau)$ as well as ${\bf s}_i(t+\tau)$ are obtained with a limited
${\cal O}(\tau^3)$ accuracy. It is worth mentioning that the interpolation
$$
[{\bf s}_i {\mbox{\boldmath $\cdot$}} {\bf s}_j]_{t+\frac{\tau}{2}}
=\frac{{\bf s}_i(t) + {\bf s}_i(t+\tau)}{2} {\mbox {\boldmath $\cdot$}}
\frac{{\bf s}_j(t) + {\bf s}_j(t+\tau)}{2}
$$
instead of Eq.~(13) is possible, in principle, too but it will not lead
to the energy conservation.

The approach considered conserves also the total linear and angular momenta,
i.e., ${\bf P}(t+\tau)={\bf P}(t)$ and ${\bf L}(t+\tau)= {\bf L}(t)$. The
first follows directly from the structure of mid-point translational forces
(8) for which $\sum_i {\bf f}_i(t+\tau/2)=0$, so that
$$
\sum_i m_i {\bf v}_i (t+\tau)=\sum_i m_i {\bf v}_i(t) .
$$
Further, taking into account Eqs.~(3) and (4), the position propagation
can be cast as
$$
{\bf r}_i(t+\tau)={\bf r}_i(t)+{\bf v}_i(t)\tau+{\bf f}_i(t+{\textstyle
\frac{\tau}{2}}) \tau^2 \big/ 2 m_i .
$$
Then the sum $\sum_i m_i {\bf r}_i(t+\tau) {\mbox{\boldmath $\times$}}
{\bf v}_i(t+\tau)$ reduces to $\sum_i m_i {\bf r}_i(t) {\mbox{\boldmath
$\times$}} {\bf v}_i(t) + \tau \sum_i {\bf r}_i(t+\tau/2) {\mbox{\boldmath
$\times$}} {\bf f}_i(t+\tau/2)$. The last term is canceled since, according
to Eq.~(8), the interparticle forces are parallel to mid-step vectors ${\bf
r}_{ij}(t+\tau/2)$, and, thus, the second property
$$
\sum_i m_i {\bf r}_i(t+\tau) {\mbox{\boldmath $\times$}} {\bf v}_i(t+
\tau)=\sum_i m_i {\bf r}_i(t) {\mbox{\boldmath $\times$}} {\bf v}_i(t)
$$
is also satisfied.

Thus, the desired algorithm of
the second order has been constructed. In view of Eqs.~(3), (4), (6)--(10),
(12) and (13), the algorithm can be presented in the following compact form
\begin{eqnarray}
&& {\bf r}_i(t+\tau) = {\bf r}_i(t) + \frac{\tau}{2} \Big[
{\bf v}_i(t)+{\bf v}_i(t+\tau) \Big] , \nonumber \\[6pt]
&& {\bf v}_i(t+\tau) = {\bf v}_i(t) - \frac{\tau}{m_i} \sum_{j (j \ne i)}
\frac{{\bf r}_{ij}(t)+{\bf r}_{ij}(t+\tau)}
{r_{ij}(t)+r_{ij}(t+\tau)} \times
\nonumber \\
&&
\bigg( \frac{\varphi(r_{ij}(t+\tau))-\varphi(r_{ij}(t))}
{r_{ij}(t+\tau)-r_{ij}(t)} -
\frac{J(r_{ij}(t+\tau))-J(r_{ij}(t))}
{r_{ij}(t+\tau)-r_{ij}(t)}
\nonumber \\[4pt]
&& \times \frac{{\bf s}_i(t) {\mbox{\boldmath $\cdot$}} {\bf s}_j(t) +
{\bf s}_i(t+\tau) {\mbox{\boldmath $\cdot$}} {\bf s}_j(t+\tau)}{2}
\bigg) , \\[5pt]
&& {\bf s}_i(t+\tau) = {\bf s}_i(t) + \frac{\tau}{\hbar}
\frac{{\bf s}_i(t)+{\bf s}_i(t+\tau)}{2} {\mbox{\boldmath $\times$}}
\nonumber \\[4pt]
&& \sum_{j (j \ne i)}
\frac{J(r_{ij}(t)) + J(r_{ij}(t+\tau))}{2}
\frac{{\bf s}_j(t) + {\bf s}_j(t+\tau)}{2} . \nonumber
\end{eqnarray}
Equation (14) constitutes, in fact, a coupled system of three nonlinear
vector equations for each particle with respect to the same number of
unknowns ${\bf r}_i(t+\tau)$, ${\bf v}_i(t+\tau)$ and ${\bf s}_i(t+\tau)$.
The system can be solved in a quite efficient way by iteration, letting
initially ${\bf v}_i^{(0)}(t+\tau)={\bf v}_i(t)$ and ${\bf s}_i^{(0)}(t+
\tau)={\bf s}_i(t)$ in the right-hand sides of Eq.~(14). Then the current
values for ${\bf r}_i(t+\tau)$, ${\bf v}_i(t+\tau)$ and ${\bf s}_i(t+\tau)$
obtained in the left-hand sides of Eq.~(14) are treated as initial guesses
for the next iteration. Already two iterations are sufficient to reach
the ${\cal O}(\tau^3)$ accuracy for the microscopic solutions and energy
conservation, i.e., $E(t+\tau)-E(t)={\cal O}(\tau^3)$. The goal of carrying
out further several updates of Eq.~(14) is to reduce the uncertainty
$\varepsilon = E(t+\tau)-E(t)$ in energy deviation to a negligibly small
value (by adjusting the number $l \ge 2$ of iterations for a given $\tau$).
The rapid convergence $\varepsilon \to +0$ is guaranteed by the relative
smallness of the step size $\tau$ and an exponential decaying of
$\varepsilon$ with increasing $l$ (see the next section).

It is interesting to remark that the total linear and angular momenta
are conserved exactly within each iteration of Eq.~(14). The reason is
that the interparticle forces are evaluated exploiting Newton's third
law and the velocities ${\bf v}_i(t+\tau)$ are updated before all ($i=
1,2,\ldots,N$) the advanced positions ${\bf r}_i(t+\tau)$ were calculated.
For similar reasons, the magnetization conservation is also fulfilled
for each iteration, when the spins are updated according to the third line
of Eq.~(14). In this case, however, the individual spin lengths will only
be preserved like energy in an iterative sense, i.e., ${\bf s}_i(t+\tau)-
{\bf s}_i(t)={\cal O}(\varepsilon)$. Nevertheless, the spin lengths can
be maintained exactly within each iteration by replacing this third line
by its mathematically equivalent counterpart (11) (were $[{\bf g}_i]_{t+
\tau/2}$ being evaluated with the help of Eq.~(12)).

\section{Numerical tests. Comparison with other methods}

In our MD simulations of the Heisenberg spin fluid (Eq.~(1)), the Yukawa
function \cite{Nijmei}
\begin{equation}
Y(r) = w \frac{\sigma}{r} \exp \left(\frac{\sigma-r}{\sigma} \right)
\end{equation}
was used to describe the spin-spin interactions. The liquid subsystem was
modelled by a soft-core potential \cite{Allen},
\begin{equation}
\varphi(r) = 4 u \left[ \left( \frac{\sigma}{r} \right)^{12} -
                 \left( \frac{\sigma}{r} \right)^6 \right] + u
\end{equation}
which accepts nonzero values at $r < 2^{1/6} \sigma$ and $\varphi(r)=0$
at $r \ge 2^{1/6} \sigma$. Here $\sigma$ is the diameter of particles,
and $u$ as well as $w$ denote the intensities of core-core and spin-spin
interactions, respectively. The simulations were carried out in a
microcanonical ensemble for $N=1000$ identical ($m_i \equiv m$, $s_i \equiv
1$) particles in a cubic box of volume $V=L^3$ employing periodic boundary
conditions. The Yukawa function was truncated at $R_{\rm c}=2.5 \sigma <
L/2$ and shifted to be zero at the truncation point to avoid the force
singularities, i.e., $J(r) = Y(r) - Y(R_{\rm c})$ at $r < R_{\rm c}$
and $J(r)=0$ otherwise. We have chosen the same thermodynamic point
as considered in previous papers \cite{Omfec,Omfes}, namely, a
reduced density of $n^\ast=N \sigma^3/V=0.6$, a reduced temperature of
$T^\ast=k_{\rm B} T/w=1.5 < T_{\rm c}^\ast$ (where $k_{\rm B}$ is the
Boltzmann's constant and $T_{\rm c}^\ast \approx 2.055$ the critical
temperature of the system \cite{Folk}), a non-zero magnetization per
particle of $|{\bf M}|/N =0.6536$ as well as the same values for the
reduced core intensity $u/w=1$ and the dynamical coupling parameter
$d=\sigma (m w)^{1/2}/\hbar=2$.

The equations of motion were solved using a well established Adams-%
Bashforth-Moulton (ABM) predictor-corrector integrator of the fourth
order \cite{Burden}, the explicit decomposition schemes \cite{Omfec,Omfes}
of the second (ED) and forth (ED4) orders, as well as our conservative
spin fluid dynamics (CSFD) algorithm (Eq.~(14)). All the test runs were
started from an identical well equilibrated configuration. A typical
example for the reduced total energy $E^\ast=E/w$ and magnetization $M=
|{\bf M}|$ per particle as depending on the length of the simulation is
shown in subsets (a) and (b) of Fig.~1, respectively, at a reduced time
step of $\tau^\ast= \tau (w/m\sigma^2)^{1/2} =0.01$.

The huge systematic deviations in the total energy obtained within the ABM
approach (see the dashed curve in Fig.~1 (a)) points out clearly that it
is highly unstable and, thus, not suitable for long-duration observations
over the system at the time step considered. We mention that in the ABM
scheme, the dynamical variables are first predicted,
\begin{eqnarray*}
{\mbox{\boldmath $\rho$}}(t+\tau)&=&{\mbox{\boldmath $\rho$}}(t)+\big[55
{\mbox{\boldmath $\dot \rho$}}(t) - 59 {\mbox{\boldmath $\dot \rho$}}(t-
\tau) \\ &+&37{\mbox{\boldmath $\dot \rho$}}(t-2\tau)-9 {\mbox{\boldmath
$\dot \rho$}}(t-3\tau)\big] \frac{\tau}{24}+{\cal O}(\tau^5) ,
\end{eqnarray*}
and further iteratively corrected as
\begin{eqnarray*}
{\mbox{\boldmath $\rho$}}(t+\tau)&=&{\mbox{\boldmath $\rho$}}(t)+\big[9
{\mbox{\boldmath $\dot \rho$}}(t+\tau)+19 {\mbox{\boldmath $\dot \rho$}}(t)
\\ &-& 5{\mbox{\boldmath $\dot \rho$}}(t-\tau)+{\mbox{\boldmath $\dot
\rho$}}(t-2\tau)\big] \frac{\tau}{24}+{\cal O}(\tau^5) ,
\end{eqnarray*}
where ${\mbox {\boldmath $\dot \rho$}}= \{ {\bf v}_i, {\bf f}_i/m, [{\bf
s}_i {\mbox{\boldmath $\times$}} {\bf g}_i]/\hbar \}$. The strong
instability of the ABM integrator can be explained by the facts that
it destroys the unit norm of spin lengths (although conserves the
magnetization vector) and generates time irreversible solutions (as is
now rigorously proved \cite{Frenkel}, the numerical stability follows
directly from the reversibility of an algorithm). For this reason, the
ABM as well as other existing predictor-corrector schemes can be used only
at very small time steps (namely, at $\tau^\ast \le 0.00125$, see Refs.
\cite{Omfec,Omfes}), were they exhibit similar equivalence in the energy
conservation as that of the decomposition algorithms. However, such small
step sizes are inefficient, because then too much expensive force and field
recalculations have to be performed in order to cover the fixed observation
time. Moreover, the ABM scheme is about twice times slower compared to the
ED integrator even if one iteration only is applied within the corrector
procedure.

No drift in the functions $E(t)$ and $M(t)$ was recognized within both
the decomposition ED and ED4 algorithms at time steps up to $\tau^\ast
\equiv \tau_{\rm max}^\ast=0.01$ and over a number of time steps of
$t/\tau= 100\,000$. These algorithms, however, do not conserve the total
energy and magnetization exactly. Instead, the last functions fluctuate
quite visibly especially in the ED case, as can be seen from Fig.~1. The
ED4 energy fluctuations are approximately in factor two smaller than
those of the ED algorithm. This compensates to some extent the additional
processor time needed to evaluate high-order expressions. But the ED4
algorithm allows to reduce the magnetization deviations to a negligible
small level which does not exceed about $\langle ({\bf M}(t)-{\bf M}(0))^2
\rangle^{1/2}/N \sim 3 \cdot 10^{-9}$ even at the greatest time step
$\tau^\ast_{\rm max}$, where $\langle \, \rangle$ denotes the microcanonical
averaging. It is worth mentioning that the ED/ED4 energy-magnetization
fluctuations are caused by the ${\cal O}(\tau^3)/{\cal O}(\tau^5)$
truncation errors and thus they will increase drastically with
increasing $\tau$.

\begin{figure}[htbp]
\begin{centering}
\begin{picture}(81,242)
\epsfxsize=81mm
\put(0,0){
\framebox{
\epsffile[69 220 540 702]{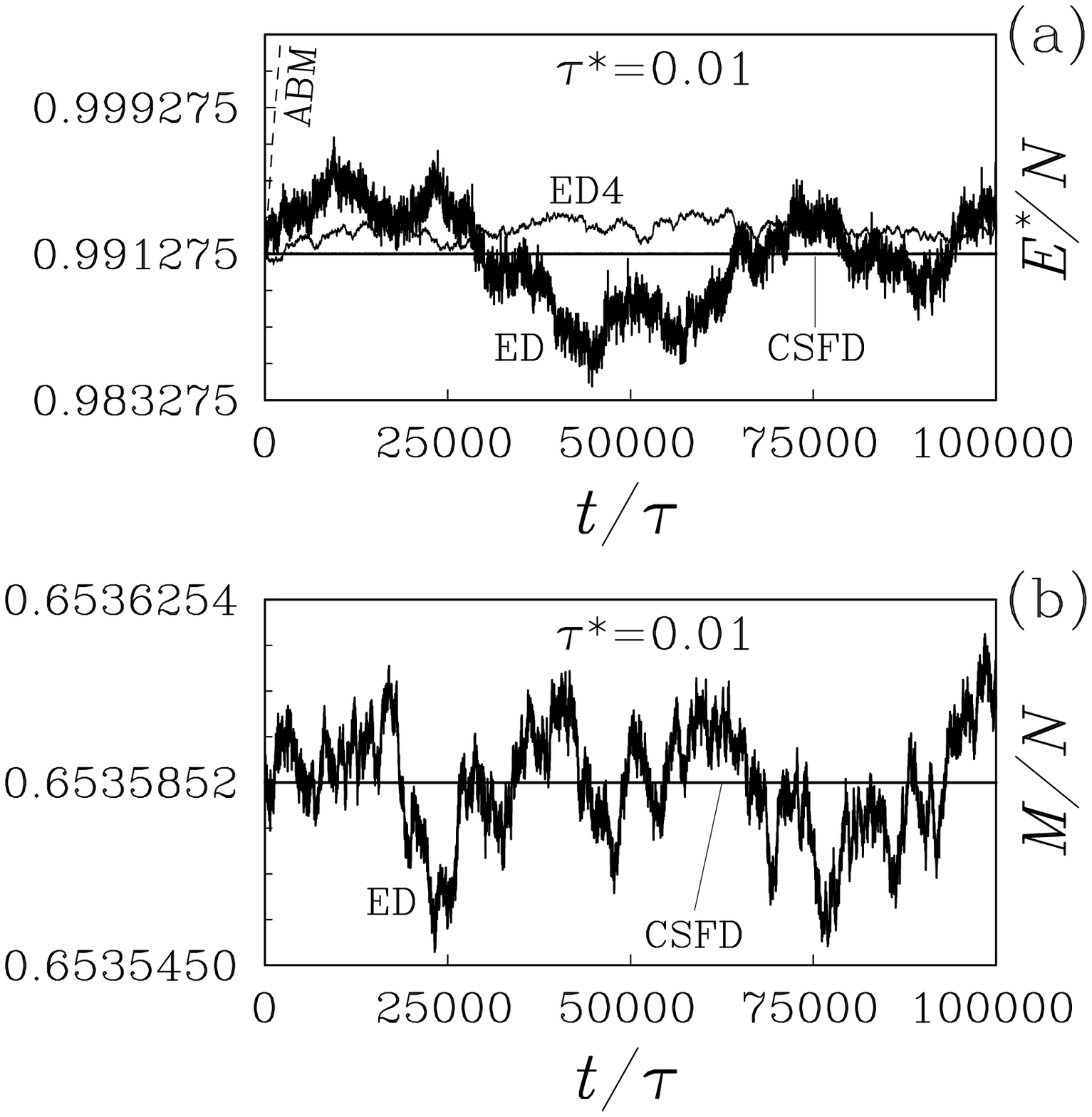}}}
\end{picture}
\end{centering}
\end{figure}

{\small
FIG.~1. The total energy $E^\ast/N$ (subset (a)) and magnetization
$M/N$ (subset (b)) per particle as functions of the length $t/\tau$
of the simulations performed for a Heisenberg spin fluid using the
predictor-corrector (dashed curve in (a), marked as ABM), decomposition
(solid curves, ED/ED4), and our new (bold solid horizontal lines, CSFD)
algorithms. Note that the ABM and ED4 curves are indistinguishable
in (b) from the CSFD line.}

\vspace{9pt}

The situation is completely different in the case of our new approach,
because the CSFD algorithm preserves the integrals of motion for arbitrary
time steps. Of course, we cannot apply too large step sizes ($\tau^\ast
\sim 1$) since then the microscopic solutions will deviate considerably
from their exact counterparts and because of too large number of iterations
needed to achieve the convergence. Choosing, for instance, $\tau^\ast \equiv
\tau^\ast_{\rm max}=0.01$ we have determined the following levels ${\cal E}$
in the averaged total energy fluctuations ${\cal E} = \langle (E^\ast(t)-
E^\ast(0))^2 \rangle^{1/2}/N$ at the end of the $100\,000$ time step runs:
$9.2 \cdot 10^{-4}$, $2.3 \cdot 10^{-5}$, $3.1 \cdot 10^{-6}$, $2.2 \cdot
10^{-7}$, and $2.8 \cdot 10^{-8}$ corresponding to the numbers $l$ of
iterations: 2, 3, 4, 6, and 8, respectively. We see, therefore, that the
iterations converge rapidly with increasing $l$ and the uncertainties can
be approximately described by the exponential dependence ${\cal E} \sim
3 \cdot 10^{-4} \exp(-1.2 l)$ at $l \ge 4$. Of course, the iterative
solutions require additional computational efforts, but they are justified
when a high level of the energy conservation is necessary. In order to
demonstrate this, we have tried to reduce the energy fluctuations within
the ED/ED4 algorithms by decreasing the time step. The corresponding result
for ${\cal E}$ at the time steps 0.01, 0.005, 0.0025, and 0.00125, i.e.,
at $\tau^\ast \equiv \tau^\ast_{\rm max}/l$ with $l=1, 2, 4$, and 8 is
presented in Fig.~2 in comparison with the above CSFD data.

\begin{figure}[htbp]
\begin{centering}
\begin{picture}(81,218)
\epsfxsize=81mm
\put(0,0){
\framebox{
\epsffile[69 499 290 702]{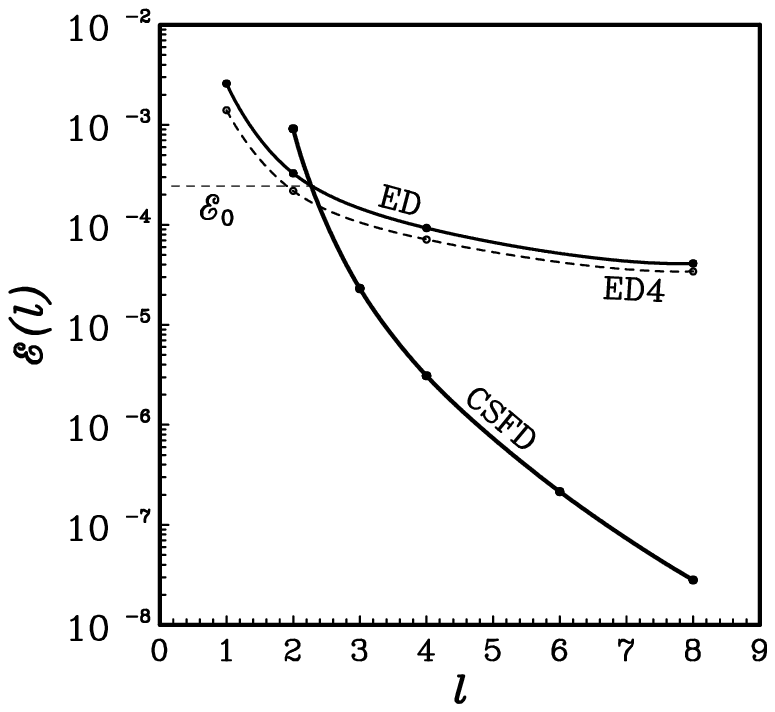}}}
\end{picture}
\end{centering}
\end{figure}

{\small
FIG.~2. The averaged total energy fluctuations ${\cal E}$ as a function of
the number $l$ of iterations, obtained in the Heisenberg fluid simulations
within the CSFD algorithm at the reduced time step $\tau^\ast_{\rm
max}=0.01$ (bold solid curve). The levels of ${\cal E}$
corresponding to the decomposition integrators at the time steps $\tau^\ast
\equiv \tau^\ast_{\rm max}/l$ are connected by solid (ED) and dashed (ED4)
curves.}

\vspace{9pt}

As one can see, such a reduction of the ED/ED4 energy fluctuations is not
efficient, since the deviations ${\cal E}$ behave like $\sim l^{-2}$,
i.e., decrease with increasing $l$ much more slower than the exponential
dependence obtained within the CSFD algorithm. In view of the results
of Fig.~2 and taking into account that at the same value of $l$ one
needs approximately the same processor time with both the ED and CSFD
algorithms (the ED4 integrator needs the time larger by factor 5 and is
less economical) in order to investigate the system over an identical time
interval, we come to the following conclusion: When the total energy
must be conserved up to a precision of ${\cal E}_0 \sim 10^{-4}$ (the
intersection point of the ED and CSFD curves, see the horizontal dashed
line in Fig.~2) or better, the preference should be done to the CSFD
algorithm. For example, a level of ${\cal E} \sim 10^{-6}$ in the
conservation is achieved at $l \sim 5$ within the CSFD algorithm, while
up at $l \sim 50$ for the ED scheme (the last value was obtained by
extrapolating the $\sim l^{-2}$ dependence to larger $l$). Thus the CSFD
algorithm appears to be approximately in 10 times faster than the ED
integrator at this level of ${\cal E}$. For ${\cal E} > {\cal E}_0$, we
can restrict ourselves to the usual explicit decomposition integrators.

Note that despite the uncertainties ${\cal E}_0 \sim 10^{-4}$ look quite
small, they can influence considerably on some observable macroscopic
quantities. The influence can be estimated quantitatively in terms of
the ratio $\Gamma=(\langle (E(t)-E(0))^2 \rangle/\langle (U(t)-U(0))^2
\rangle)^{1/2}$ of total and potential energy fluctuations. For our
system ${\cal U}=\langle (U^\ast(t)-U^\ast(0))^2 \rangle^{1/2}/N \sim
10^{-2}$, where $U^\ast=U/w$, so that $\Gamma_0={\cal E}_0/{\cal U} \sim
1\%$. Usual investigated quantities, such as thermodynamic functions,
structure factors, etc., will be calculated approximately with the same
relative precision $\Gamma_0$ (provided the averaging over the produced
trajectories is performed during a sufficiently large time interval to
be entitled to ignore the statistical noise). But when long-tail time
correlation functions or derivatives of the thermodynamic functions are
involved into the computations, the impact of the artificial energy
fluctuations on the results will be much greater. For instance, the
relative uncertainty in the measurements of the specific heat (which
are based on a microcanonical ensemble fluctuation formula) is estimated
to be already $(\Gamma_0)^{1/2} \sim 10\%$. This uncertainty may appear
to be too large to determine correctly a phase diagram of the system.

A similar pattern to that shown in Fig.~2 was observed within the CSFD
approach at greater time steps $\tau > 0.01$. The energy as well as
magnetization fluctuations continued to damp exponentially with increasing
$l$, although a greater number of the iterations was necessary to reach
the same level of the conservation. Note that a rotational matrix version
of the CSFD algorithm (when the third line of Eq.~(14) is replaced by
Eq.~(11)), in which individual spin lengths are maintained exactly within
each iteration, leads to a somewhat better energy preservation at a given
$l$ (but then the total magnetization like energy will be conserved in
the iterative sense, i.e., at sufficiently large $l$). The CSFD results
presented above for ${\cal E}$ have been obtained using this version
for spin subdynamics propagations.

Further improvements in the efficiency of the CSFD algorithm can be reached
applying the following computational trick. It can occur that after a some
period of time during the integration process the energy difference $E(t)-
E(0)$ corresponding to the last $l$-th iteration (within a current time
step $t/\tau$) exceeds the difference obtained for the previous $(l-1)$-th
iteration. Such a situation is possible because of round-off errors and
an accumulation of other numerical uncertainties, especially at relatively
small values of $l$, where the lack in the time-reversibility can lead to
an instability of the solutions (note the CSFD algorithm is time-reversible
in the iterative sense, i.e., at large enough values of $l$). Then to avoid
the accumulation, we should take merely the values for microscopic phase
variables corresponding to this previous $(l-1)$-th iteration. The trick
with a flexible number of the iterations will guarantee a good stability
for small $l \sim$ 2--4 as well.

Another technical detail concerns the way in which the expression $[\xi(r(t+
\tau))-\xi(r(t))]/[r(t+\tau)-r(t)]$ (appearing in Eq.~(14) for $\xi \equiv
\varphi,J$) should be treated in the limit $r(t+\tau) \to r(t)$. As was
pointed out earlier, this expression must be computed using its limiting
representation $\xi'([r(t)+r(t+\tau)]/2)+\epsilon^2 {\cal O}(\tau^2)$
when the difference $|r(t+\tau) - r(t)| < \epsilon$ is small enough.
Then letting $\epsilon^2$ being equal a machine zero, the truncated term
$\epsilon^2 {\cal O}(\tau^2)$ can be ignored completely. In our program
code we have used a double precision throughout with 16 significant digits,
$\epsilon^2 = 10^{-16}$, so that the value $\epsilon$ was set to be equal
to $10^{-8}$. It is interesting to remark that the condition $|r_{ij}(t+\tau)
-r_{ij}(t)| < \epsilon$ was never achieved for any pair $ij$ of particles
during the simulations and, thus, the limiting expression was never used.
This can be explained by the fact that the probability of finding the system
in such a state is prohibitively small and is proportional to $C \epsilon$.
The coefficient $C$ increases with increasing the length of the simulations
and the number of particles as $C \sim t N^2$. Thus, the limiting expression
is expected to be applied for systems with a greater size or when extra long
simulations are performed.

Finally, some words about the angular momentum conservation. As is well
known, the periodic boundary conditions, which are commonly used in MD
simulations to reduce the finite-size effects, destroy the angular momentum
vector. Nevertheless, it has been established that this vector is conserved
in our simulations in mean, namely, $\langle {\bf L}(t) \rangle \approx
{\bf L}(0)$. Note that initial values for the total angular as well linear
momenta were putted to be equal to zero, ${\bf L}(0)=0$ and ${\bf P}(0)=0$,
i.e., the system was considered at the very beginning as one which does
not move as a whole translationally and rotationally.

\section{Applications to other systems: pure liquids
         and harmonic oscillator}

The algorithm derived in the preceding section can also be
applied with equal successes to dynamics simulations of other liquid models.
For instance, letting formally $J \equiv 0$, we come to the usual equations
of motion corresponding to a pure liquid system. These equations can be
integrated using the first two lines of the same propagation equation (14),
where the terms with $J$ in the right-hand side of the second line must be
omitted (the third line describes spin subdynamics and is not relevant in
this case).

Our simulations of pure liquid dynamics were based on a system composed
of $N=256$ particles interacting through a cut-off Lennard-Jones (LJ)
potential, $\varphi(r)=\Phi(r)-\Phi(R_{\rm c})$ at $r<R_{\rm c}=3.25
\sigma$ with $\varphi(r)=0$ otherwise, where
\begin{equation}
\Phi= 4 u \left[ \left( \frac{\sigma}{r} \right)^{12} -
          \left( \frac{\sigma}{r} \right)^6 \right] .
\end{equation}
The MD test runs have been performed at a reduced density of $n^\ast=0.845$
and a reduced temperature of $T^\ast=k_{\rm B}T/u=1.7$. For the purpose
of comparison, the equations of motion were integrated applying also a
well-established velocity Verlet (VV) algorithm \cite{Swope,Frenkel} of
the second order and its forth-order (VV4) counterpart \cite{Suzuki}. Our
algorithm we will call now as CPFD (conservative pure fluid dynamics). A
typical maximal value for the reduced time step in simulating such a system
is $\tau^\ast_{\rm max}=\tau (u/m\sigma^2)^{1/2} =0.005$ \cite{Omelf}. All
the runs started from a well equilibrated configuration and covered an
identical time interval of $t^\ast=t (u/m\sigma^2)^{1/2}=50$ (corresponding
to $10\,000$ time steps at $\tau^\ast=0.005$).

It is worth mentioning that the explicit VV integrator propagates the phase
variables according to the relations
\begin{eqnarray*}
{\bf r}_i(t+\tau)&=&{\bf r}_i(t)+{\bf v}_i(t) \tau +
{\bf f}_i(t) \frac{\tau^2}{2m}+{\cal O}(\tau^3) , \\
{\bf v}_i(t+\tau)&=&{\bf v}_i(t)+\big[{\bf f}_i(t)+
{\bf f}_i(t+\tau)\big] \frac{\tau}{2m}+{\cal O}(\tau^3) .
\end{eqnarray*}
This propagation can be presented as
$$
\big\{ {\bf r}_i(t+\tau), {\bf v}_i(t+\tau) \big\} = {\bf D}(t,\tau)
\big\{ {\bf r}_i(t), {\bf v}_i(t) \big\}+{\cal O}(\tau^3) ,
$$
where ${\bf D}(t,\tau)$ denotes the evolutionary operator. The VV4 algorithm
deals (similarly to the ED4 scheme) with the five stages propagation
$$
\big\{ {\bf r}_i(t+\tau), {\bf v}_i(t+\tau) \big\}=\prod_{k=1}^5
{\bf D}(t,\xi_k \tau) \big\{ {\bf r}_i(t), {\bf v}_i(t) \big\}+
{\cal O}(\tau^5) ,
$$
where the coefficients $\xi_k$ are: $\xi_1 = \xi_2 = \xi_4 = \xi_5 \equiv
\xi = 1/(4 - 4^{1/3})$, and $\xi_3 = 1 - 4\xi$. The VV approach needs only
in one force evaluation (the most time-consuming part of the calculations)
per time step, $p_{\rm VV}=1$, while $p_{\rm VV4}=5$. The CPFD algorithm
requires two force evaluation per iteration within the time step,
$p_{\rm CPFD}=2$.

The averaged total energy fluctuations ${\cal E}=\langle (E(t)-E(0))^2
\rangle^{1/2}/(uN)$ obtained within the CPFD integration at the time step
$\tau^\ast_{\rm max}=0.005$ and the numbers of iterations of $l=2, 3, 4$,
and 8 are plotted in Fig.~3 as a function of the reduced processor time
$l p$ (where in this case $p=p_{\rm CPFD}$) needed to perform the run
of the mentioned above length $t^\ast=50$. The fluctuations identified
during the integration at the time steps $\tau^\ast=\tau^\ast_{\rm max}
/l$ using the VV algorithm with $l=1,2,4,8$ and 16 as well as the VV4
algorithm with $l=0.5,1,2,4$ are also included in this figure. The
processor time spent to carry out the VV run of the length $t^\ast=50$
at $\tau^\ast=0.005$ is assumed to be equal to unity in our dimensionless
presentation $l p$ (where $p=1,2$ and 5 for the VV, CPFD, and VV4
integrators).

The LJ energy fluctuations damp with increasing $l$ like $\sim l^{-2}$,
$\sim l^{-4}$, and $\sim \exp(-2.4 l)$ within the VV, VV4, and CPFD
integrations, respectively. Up to three intersection points corresponding
to the VV--VV4, VV4--CPFD, and VV4--CPFD curves with the energy conservation
levels of ${\cal E}_1 \sim 6 \cdot 10^{-5}$, ${\cal E}_2 \sim 10^{-5}$, and
${\cal E}_3 \sim 3 \cdot 10^{-7}$ can be observed in Fig.~3. So that the
usual VV algorithm is recommended to be used when the precision ${\cal E}$
of energy conservation plays not so important role in the computations,
namely, when ${\cal E} \ge {\cal E}_1$. The calculation with the help of
the VV4 integrator appears to be most computationally efficient in the
intermediate regime ${\cal E}_3 < {\cal E} < {\cal E}_1$. Finally, when a
very accurate conservation, ${\cal E} < {\cal E}_3$, is required, the best
choice is to apply the CPFD algorithm because then it becomes to be most
economical.

\begin{figure}[htbp]
\begin{centering}
\begin{picture}(81,218)
\epsfxsize=81mm
\put(0,0){
\framebox{
\epsffile[69 497 291 702]{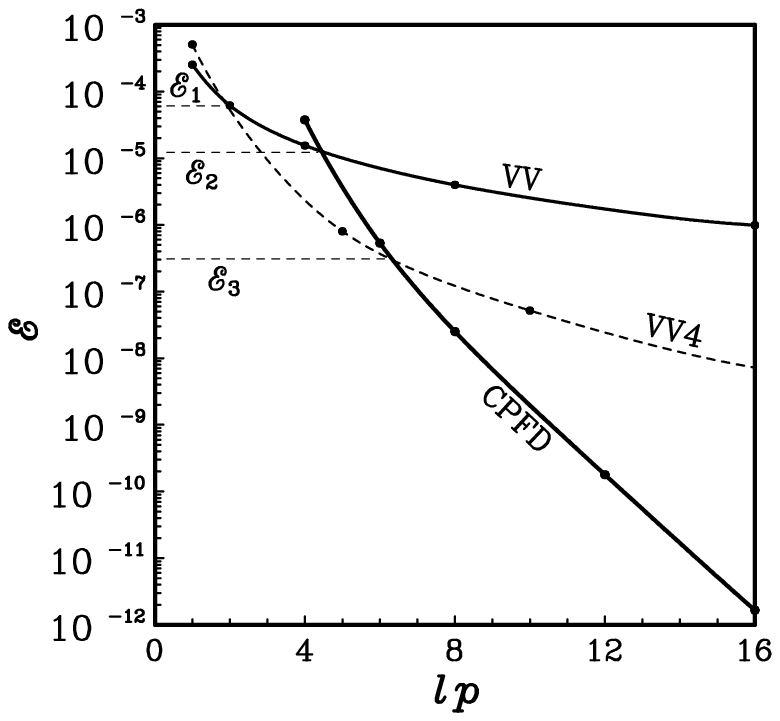}}}
\end{picture}
\end{centering}
\end{figure}

{\small
FIG.~3. The averaged total energy fluctuations ${\cal E}$ as a function
of the reduced processor time $l p$ needed for the simulations of a
Lennard-Jones liquid within the CPFD (bold solid curve), usual velocity
Verlet (solid curve, VV), and fourth-order
velocity Verlet (dashed curve, VV4) algorithms.}

\vspace{9pt}

The CPFD approach can also be used for the prediction of dynamical phenomena
in other many-body collections (such as the solar system, for instance)
and treated as an efficient numerical solver of first-order differential
equations. The most notorious example (which allows to be analyzed
analytically) is the equation ${\rm d}^2 x/{\rm d} t^2 \equiv \ddot
x = -x$ describing dynamics of a simple harmonic oscillator. This equation
reduces to a system of two first-order differential equations, $\dot x=v$
and $\dot v=-x$, which in turn can be reproduced from the first two lines
of general equation (2) putting formally ${\bf r}_i \equiv x$, ${\bf v}_i
\equiv v$, $\varphi=x^2/2$, $m_i \equiv 1$ and $J=0$. Then in view of
Eq.~(14), the time propagation reads $x(t+\tau)=x(t)+\tau [v(t)+v(t+
\tau)]/2$ and $v(t+\tau)=v(t)-\tau [x(t)+x(t+\tau)]$. The last two
relations can be solved explicitly, and the result for the conservative
numerical trajectories is
\begin{eqnarray*}
x(t+\tau)&=&\frac{x(t) \big(1-\tau^2/4 \big) +
v(t)\tau}{1+\tau^2/4} , \\ [4pt]
v(t+\tau)&=&v(t)-\tau\frac{x(t)+v(t) \tau/2}{1+\tau^2/4} ,
\end{eqnarray*}
whereas the VV solutions are
\begin{eqnarray*}
x(t+\tau)&=&x(t) \big(1-\tau^2/2\big)+v(t)\tau , \\ [4pt]
v(t+\tau)&=&v(t) - \tau \big[x(t) \big(1-\tau^2/4\big)+v(t)\tau/2\big] .
\end{eqnarray*}

Choosing the initial conditions $x(0)=0$ and $\dot x(0) \equiv v(0)=1$,
the above two types of numerical trajectories can be compared between
themselves and with respect to the exact solution $x(t)=\sin(t)$ and
$\dot x(t) \equiv v(t)=\cos(t)$. The result of comparison for $x(t)$
is presented in Fig.~4 at a typical time step of $\tau = 0.05 T$, where
$T=2\pi$ denotes the period of the oscillations. As can be seen easily,
the conservative solution leads to a better reproduction of the original
dependence than the VV trajectory, despite the both CPFD and VV approaches
are valid to the same ${\cal O}(\tau^3)$ order in truncation errors.
Therefore, additional cancellations of the truncation uncertainties are
possible due to the exact preservation of the integral of motion $2E=
\dot x^2 + x^2 \equiv 1$ along the CPFD trajectory (note that maximal
VV deviations in $E$ consist about 20\% at $\tau = 0.05 T$). Similar
cancellations of the truncation uncertainties in microscopic solutions
within our conservative approach should be expected for other systems
of differential equations, in particular, for spin and pure liquid
dynamics.

\begin{figure}[htbp]
\begin{centering}
\begin{picture}(81,218)
\epsfxsize=81mm
\put(0,0){
\framebox{
\epsffile[69 501 288 702]{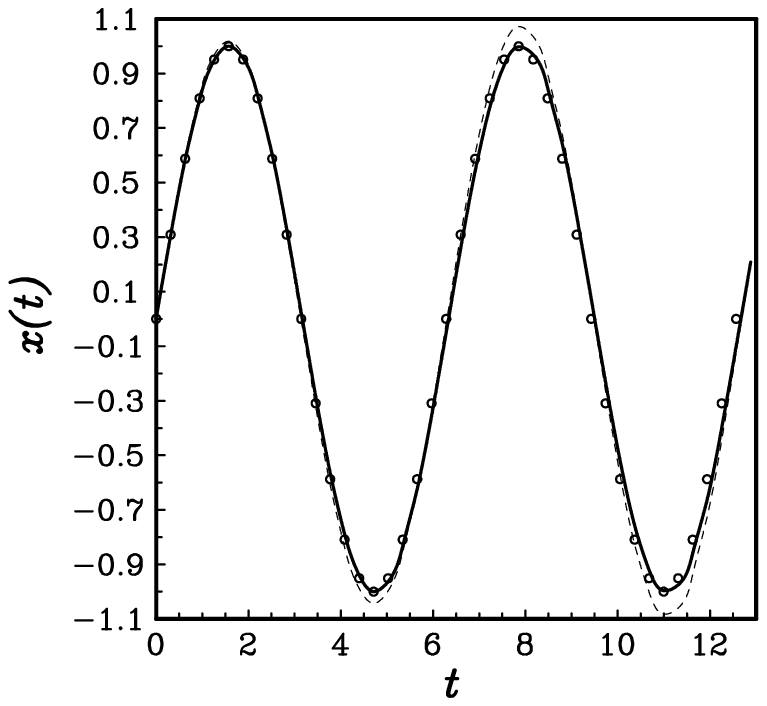}}}
\end{picture}
\end{centering}
\end{figure}

{\small
FIG.~4. Numerical solutions to the differential equation $\ddot x(t)=-x$
obtained during our new (solid curve) and velocity-Verlet (dashed
curve) integrations at the time step $\tau = 0.05 T$ with $T=4\pi$. The
exact result $x(t)=\sin(x)$ is shown as open circles.}

\section{Concluding remarks}

One of the most fundamental characteristics in physics are the conservation
laws. Therefore, it is desirable that the numerical methods in computational
physics obey these laws. Unfortunately, the most popular algorithms,
such as predictor-corrector, Runge-Kutta, Verlet, decomposition Suzuki-%
Trotter, etc., being applied to the nonlinear many-body problem, do not
preserve fundamental physical invariants, like energy and angular momentum,
when these are inherent in the description.

In the present paper we have tried to remedy such a situation and
formulated a novel completely conservative approach for numerical
integration of the equations of motion in classical systems. The
approach is general enough to be used for a wide class of systems
such as spin and pure liquids, collections of charged particles, etc.
It can also be considered for the prediction of other phenomena in
physics, astrophysics, chemistry and biology, whenever the numerical
solutions to systems of differential equations are necessary.

Our main attention in this study was concentrated on dynamics of spin
liquid models in which additional effects with respect to pure liquids
are possible because of the energy exchange between spin and liquid
subsystems \cite{MryglodFolk,Omfec,Omfes}. As a result, a new second-order
MD algorithm (called as CSFD) has been consequently derived within the
above presented approach. Its greatest advantage is that all the
integrals of motion existing in the system, namely, the total energy,
linear and angular momenta, individual spin lengths and total magnetization
are conserved independently on the size of the time step. It is worth
emphasizing that such a complete conservation has been achieved
intrinsically, i.e., without the introduction of any artificial external
forces or numerical constraints. Moreover, the resulting algorithm
maintains the time reversibility property inherent in the basic
equations. This is also important for long-duration MD observations
because the stability of an algorithm is closely connected with
its time reversibility \cite{Frenkel}.

The presented algorithm is implicit, i.e., it requires iterative solutions.
Thus, when a high precision in conservation is not needed, the CSFD scheme
may be less efficient in practice than explicit decomposition methods
\cite{Omfec,Omfes}. We have shown, however, on the basis of actual
simulation of a Heisenberg fluid model that when the total energy and
magnetization must be reproduced precisely, the CSFD algorithm
may be in order or even more faster than the decomposition integrators.
Another important feature of the conservative method is that additional
cancellations of the truncation uncertainties are possible in microscopic
solutions due to the exact preservation of the macroscopically observable
integrals of motion, as was demonstrated analytically on a simple example
of the harmonic oscillator.

For a particular case when the spin subsystem is absent, the CSFD
algorithm reduces to a so-called CPFD integrator. While this work has
been done we have learned that this integrator is equivalent, in fact,
to that developed independently by Greenspan \cite{Green} as well as
Gonzalez and Simo \cite{Gonza}. These authors, however, considered the
integration in respect of applying it to mechanical systems when the
number of particles is not very large. Here we have shown within the
LJ model that the CPFD integrator can be used with equal success in
MD simulations of pure liquids.

The approach presented can be adapted to many-component systems,
optimized to a multiple time stepping integration and extended to
higher-order versions. These and other related problems will be the
subject of a separate investigation.

\section{Acknowledgements}

Part of this work was supported by the Fonds zur F\"orderung
der wissenschaftlichen Forschung under Project No. P12422-TPH.

\end{multicols}


\begin{thebibliography}{26}

\bibitem{ChenLan}
 K. Chen and D.P. Landau, Phys. Rev. B {\bf 49}, 3266 (1994).

\bibitem{BunChen}
 A. Bunker, K. Chen, and D.P. Landau, Phys. Rev. B {\bf 54}, 9259
 (1996).

\bibitem{EvLan}
 H.G. Evertz and D.P. Landau, Phys. Rev. B {\bf 54}, 12302 (1996).

\bibitem{CosLan}
 B.V. Costa, J.E.R. Costa, and D.P. Landau, J. Appl. Phys. {\bf 81},
 5746 (1997).

\bibitem{Krech}
 M. Krech, A. Bunker, and D.P. Landau, Comp. Phys. Commun. {\bf 111},
 1 (1998).

\bibitem{LanKre}
 D.P. Landau and M. Krech, J. Phys. Cond. Mat. {\bf 11}, R179 (1999).

\bibitem{Landau}
 D.P. Landau, S.-H. Tsai, M. Krech, and A. Bunker, Int. J. Mod. Phys.
 C {\bf 10}, 1541 (1999).

\bibitem{Tsai}
 S.-H. Tsai, A. Bunker, and D.P. Landau, Phys. Rev. B {\bf 61},
 333 (2000).

\bibitem{Omfe}
 I.P. Omelyan, I.M. Mryglod, and R. Folk, Europhys. Lett., to be published.

\bibitem{Lom}
 E. Lomba, J.J. Weis, N.G. Almarza, F. Bresme, and G.
 Stell, Phys. Rev. E {\bf 49}, 5169 (1994).

\bibitem{Tavar}
 J.M. Tavares, M.M.T. Gama, P.I.C. Teixeira, J.J. Weis, and
 M.J.P. Nijmeijer, Phys. Rev. E {\bf 52}, 1915 (1995).

\bibitem{Mryglod}
 I.M. Mryglod, M.V. Tokarchuk, and R. Folk, Physica A {\bf 220},
 325 (1995).

\bibitem{Nijmeis}
 M.J.P. Nijmeijer, J.J. Weis, Phys. Rev. Lett. {\bf 75}, 2887 (1995).

\bibitem{Nijmei}
 M.J.P. Nijmeijer, J.J. Weis, Phys. Rev. E {\bf 53}, 591 (1996).

\bibitem{MryglodFolk}
 I. Mryglod, R. Folk, S. Dubyk, and Yu. Rudavskii, Physica A {\bf 277},
 389 (2000).

\bibitem{Omfec}
 I.P. Omelyan, I.M. Mryglod, and R. Folk, Condens. Matter Phys. {\bf 3},
 497 (2000).

\bibitem{Omfes}
 I.P. Omelyan, I.M. Mryglod, and R. Folk, Phys. Rev. Lett., to be published
 (electronic version is available at http://xxx.lanl.gov/abs/cond-mat/0012059).

\bibitem{Suzuki}
 M. Suzuki and K. Umeno in {\em Computer Simulation Studies in
 Condensed Matter Physics VI}, edited by D.P. Landau, K.K. Mon, and
 H.B. Sch\"uttler (Springer, Berlin, 1993).

\bibitem{Swope}
 W.C. Swope, H.C. Andersen, P.H. Berens, and K.R. Wilson,
 J. Chem. Phys. {\bf 76}, 637 (1982).

\bibitem{Burden}
 R.L. Burden and J.D. Faires, {\em Numerical Analysis}, 5th ed.
 (PWS Publishing, Boston, 1993).

\bibitem{Allen}
 M.P. Allen and D.J. Tildesley, {\em Computer Simulation of Liquids}
 (Clarendon, Oxford, 1987).

\bibitem{Folk}
 R. Folk, I.M. Mryglod, and I.P. Omelyan [unpublished].

\bibitem{Frenkel}
 D. Frenkel and B. Smit, {\em Understanding Molecular Simulation:
 from Algorithms to Applications} (Academic Press, New York, 1996).

\bibitem{Omelf}
 I.M. Mryglod, I.P. Omelyan, and M.V. Tokarchuk,
 Mol. Phys. {\bf 84}, 235 (1995).

\bibitem{Green}
 D. Greenspan, Computers Math. Applic. {\bf 29} No 4, 37 (1995).

\bibitem{Gonza}
 O. Gonzalez, J.C. Simo, Comput. Methods Appl. Mech. Engrg. {\bf 134},
 197 (1996).

\end{thebibliography}
\end{document}